\documentclass[%
reprint,
showpacs,preprintnumbers,
 amsmath,amssymb,
 aps,
prb,
]{revtex4-1}

\usepackage{graphicx}
\usepackage{dcolumn}
\usepackage{bm}
\usepackage{color}
\def\etal{\textit{et al.}~}

\begin{document}


\title{Effect of Dynamic Surface Polarization on the Oxidative Stability of Solvents in Nonaqueous Li-O$_2$ Batteries}

\author{Abhishek Khetan}
\affiliation{Institute for Combustion Technology, RWTH, Aachen, Germany, 52056}
\author{Heinz Pitsch}
\affiliation{Institute for Combustion Technology, RWTH, Aachen, Germany, 52056}
\author{Venkatasubramanian Viswanathan}%
\email{venkvis@cmu.edu}
\affiliation{Department of Mechanical Engineering, Carnegie Mellon University, Pittsburgh, Pennsylvania 15213}
\affiliation{Department of Physics, Carnegie Mellon University, Pittsburgh, Pennsylvania 15213}

\date{\today}

\begin{abstract}
Polarization-induced renormalization of the frontier energy levels of interacting molecules and surfaces can cause significant shifts in the excitation and transport behavior of electrons. This phenomenon is crucial in determining the oxidative stability of nonaqeous electrolytes in high energy density electrochemical systems such as the Li-O$_2$ battery. On the basis of partially self-consistent first-principles ScGW0 calculations, we systematically study how the electronic energy levels of four commonly used solvent molecules, namely dimethylsulfoxide (DMSO), dimethoxyethane (DME), tetrahydrofuran (THF) and acetonitrile (ACN), renormalize when physisorbed on the different stable surfaces of Li$_2$O$_2$, the main discharge product. Using band level alignment arguments, we propose that the difference between the solvent's highest occupied molecular orbital (HOMO) level and the surface's valence band maximum (VBM) is a refined metric of oxidative stability. This metric and a previously used descriptor, solvent's gas phase HOMO level, agree quite well for physisorbed cases on pristine surfaces where ACN is oxidatively most stable followed by DME, THF and DMSO. However, this effect is intrinsically linked to the surface chemistry of solvent's interaction with the surfaces states and defects, and depends strongly on their nature. We conclusively show that the propensity of solvent molecules to oxidize will be significantly higher on Li$_2$O$_2$ surfaces with defects as compared to pristine surfaces. This suggests that the oxidatively stability of solvent is dynamic and is a strong function of surface electronic properties. Thus, while gas phase HOMO levels could be used for preliminary solvent candidate screening, a more refined picture of solvent stability requires mapping out the solvent stability as a function of the state of the surface under the operating conditions.
\end{abstract}

\maketitle


\section{\label{sec:level1}Introduction}

It is evident from photoemission and electron transport measurements that a molecule's energy levels can be dynamically influenced by the polarizability of its local electrochemical environment~\cite{Kahn2003,Repp2005}. As the molecule moves closer to the surface, such polarization effects due to image-charge interactions can cause sizeable shifts in its frontier orbital energy levels which can drastically alter its propensity for charge transfer with the electrode surfaces. The dynamic polarization of the surface in turn depends on the charge state and orientation of the approaching solvent molecule~\cite{Garcia:2009aa, Garcia:2011aa}. These mechanisms have been shown to lead to a narrowing of the band gap at semiconductor-metal interfaces using many-body calculations~\cite{Neaton2006,Garcia:2009aa,Garcia:2011aa,Siegel2016}. Detailed studies on renormalization effects for adsorbed molecules, especially solvent molecules relevant of Li-ion batteries are very limited~\cite{Garcia:2011aa}. 

Accounting for such effects is of utmost importance, especially in the context of the nonaqueous Li-O$_2$ battery.  This battery chemistry possesses one of the highest energy densities of all beyond Li-ion (BLI) concepts.  In the nonaqueous Li-O$_2$ battery, Li$_2$O$_2$ is the main discharge product, through the overall reaction, given by~\cite{abraham1996polymer,Girishkumar:2010aa} in a reaction given by
\begin{equation} \label{eq:li2o2}
\mathrm{Li{_2}O{_2}} \leftrightharpoons \mathrm{2Li{^+} + O{_2} + 2e{^-}}
\end{equation}
These batteries are plagued by issues regarding electrolyte stability and limited charge-transport of the discharge product.\cite{oleg2015TMR} The low rechargeability of Li-O$_2$ cells is characterized by rapidly increasing overpotentials due to electrolyte degradation~\cite{alan2014ChemRev,Balaish:2014aa}.

Previous experimental and theoretical studies have emphasized the importance of oxidative stability of nonaqueous solvents as the one of the key factors determining battery's rechargeability~\cite{McCloskey:2011aa,McCloskey:2012aa,McCloskey:2013aa,Bryantsev:2012aa,Bryantsev:2013ab,Khetan:2014aa,Khetan:2014ab,Khetan:2015aa}. \textit{McCloskey}~\etal, showed using in-situ quantitative differential electrochemical mass spectrometry (DEMS) and ex-situ analysis of discharge products using XRD, NMR and Raman spectroscopy~\cite{McCloskey:2011aa,McCloskey:2012aa,McCloskey:2013aa}, that the majority of solvent degradation in Li-O$_2$ batteries occurs during the charging through a process of solvent oxidation and carbonate formation.

\begin{figure}[!ht]
\centering
\includegraphics[width=0.48\textwidth]{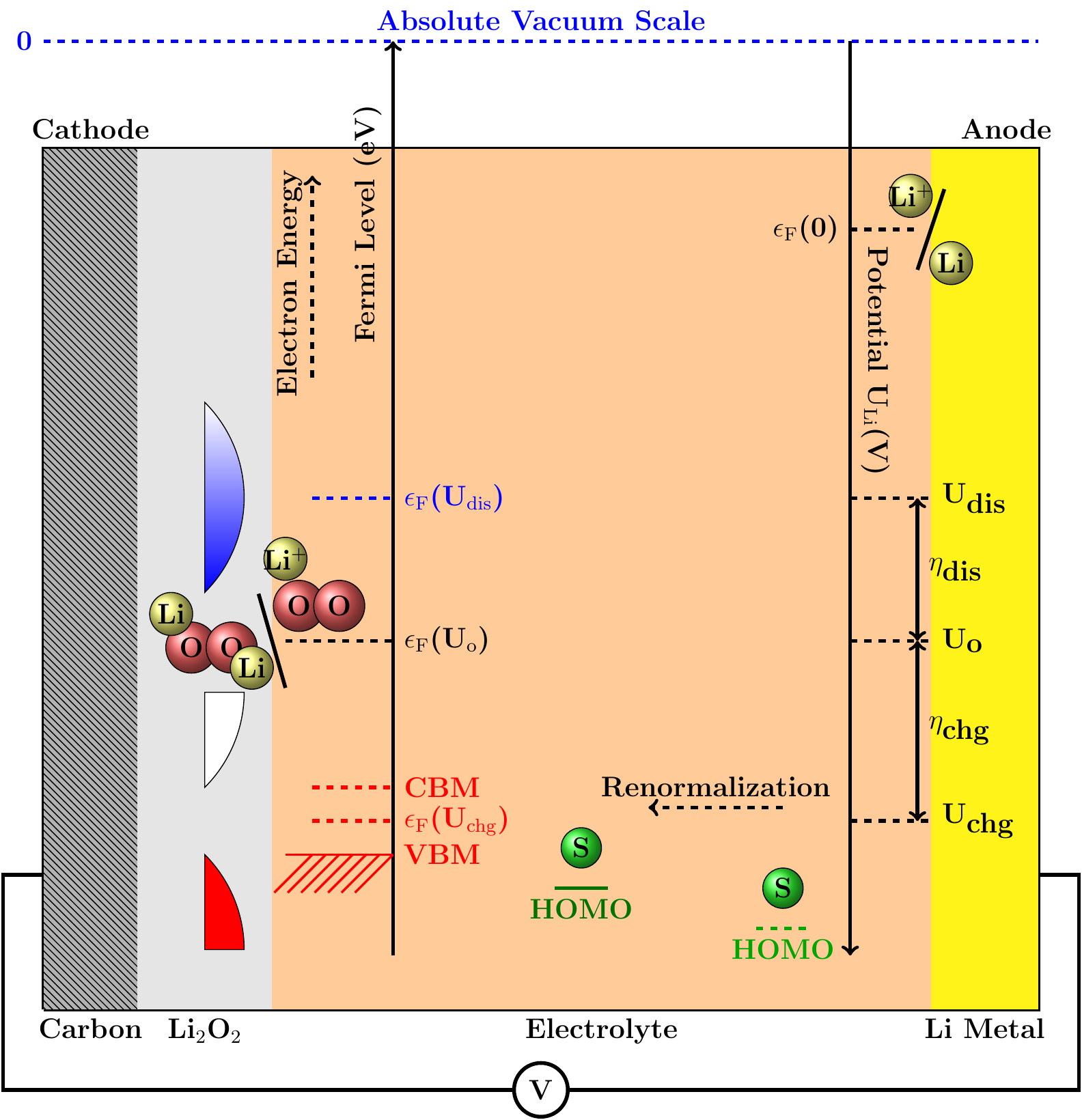}
\caption{Schematic showing the alignment of the Fermi levels, $\epsilon_{\mathrm{F}}$, at the cathode (left) with the corresponding electrode potentials U$_{\mathrm{Li}}$ versus the Li/Li$^+$ redox couple at the anode (right). The surface stable during charge is shown as semi-conducting with filled (red) and empty (white) bands, and the surface stable during discharge is shown as metallic with a continuous distribution of states (blue) at the Fermi level. The renormalization of the solvent's HOMO level to more positive value is shown upon approach to the cathode surface.}
\label{fig:new_band_scheme}
\end{figure}

In an earlier work, we have demonstrated that to a first approximation, the solvent's HOMO level is a descriptor for describing oxidative stability of the solvents~\cite{Khetan:2014aa}. However, there is a renormalization of solvent's electronic energy levels upon physisorption often leads to the shrinking of its electrochemical stability window, making it suspectible to oxidation during charge.   The extent of this effect in case of Li-O$_2$ batteries needs to be explored.   It is worth highlighting that Li$_2$O$_2$ has a large band gap and is known to be strongly insulating in the bulk~\cite{Garcia:2011ab,Radin:2012aa}. However, the electronic structure of the stable surfaces varies as a function of the electrode potential, as shown schematically in Fig.~\ref{fig:new_band_scheme}.

From a theoretical standpoint, the effective non-interacting Kohn-Sham (KS) scheme within the density functional theory (DFT) is widely know to be inadequate at describing band structures and band edge locations, especially in highly correlated semi-conducting and insulating materials~\cite{Burke2009,Garcia:2009aa,Garcia:2011aa,Garcia:2011ab}. While isolated molecules can still be described by making use of hybrid functionals, they are insufficient for capturing correlation effects due to image charge interactions between the surface and the solvent molecule. The Many-body perturbation theory provides a systematic formalism, known as the GW method, to obtain the true single-particle excitations energies, also known as the quasiparticle (QP) energies.
 
In this work, we direct our focus on systematically examining the dynamic polarization of the surface and renormalization of the solvent molecule's HOMO level using high fidelity \textit{ab initio} calculations. We consider the most stable surfaces of Li$_2$O$_2$, under charge and discharge conditions with a chosen set of nonaqueous solvents analyzed previously, namely, acetonitrile (ACN), tetrahydrofuran (THF), dimethoxyethane (DME), and dimethylsulfoxide (DMSO), as they span of over a wide range of dielectric constants, dipole moments and geometries. In order to emphasize the variance in the degree of renormalization during transient processes such as charging and discharging, we also consider the effect of two kinds of defects on the each for these terminations, namely, a defect due to a missing Li atom and due to a missing Li$_2$O$_2$ molecule. We discuss their effect on the renormalization of the electronic levels and hence the oxidative stability of solvent molecules in comparison to the pristine surfaces. 

\section{\label{sec:level1}Computational Methods}

The accurate determination of an interacting molecule's renormalized frontier energy levels using the GW method on several metallic and semiconducting surfaces has been performed in some recent studies~\cite{Garcia:2009aa,Garcia:2011aa,Neaton2006,Siegel2016}. The details of the quasiparticle GW formalism can be found in several seminal works by \textit{Hedin}~\etal~\cite{Hedin:1971a,Hedin:1971b} and others~\cite{Bruneval:2009,Kotani:2007}. In this work, the calculations were performed using the Vienna ab initio Simulation Package (VASP)~\cite{kresse1996I,kresse1996II}, in which the core-valance electron interactions were treated using the projector augmented wave (PAW) formalism~\cite{kresse1999,Blochl1994}. The partially self-consistent ScGW0 algorithm in VASP, involves full updates of the quasiparticle energies and electron orbitals performed in the calculation of G only.~\cite{Bruneval:2009,Kotani:2007}  The fully self-consistent algorithm is known to over-predict band gaps by 10-15 $\%$~\cite{Shishkin2007}. The exactly diagonalized wavefunctions resulting from the generalized gradient approximation (GGA) calculations with PBE functionals were used as inputs for applying the quasi-particle corrections. Although expensive, this method yields much more reliable estimates for band structures and locations of band edges in highly correlated materials, such as Li$_2$O$_2$, which are usually poorly described by conventional semi-local and hybrid functionals~\cite{Falco:2013aa}.

It is known from previous theoretical studies, that the $1\bar{1}00$ facet of Li$_2$O$_2$ is the most stable one under charging conditions, while the $0001$ facet is the most stable while discharging ~\cite{Hummelshoj:2013aa,Radin:2012aa,Niu2015}. Both the surface facets of the cathode surface were approximated with a slab of 3 stoichiometric Li$_2$O$_2$ layers. The slabs contained a total of 24 Li and 24 O atoms, where the atoms in the bottom-most stoichiometric layer remained fixed to their bulk positions. The total slab height was set at z = 25~\AA, which afforded a vacuum gap of 17~\AA, with a rectangular 6.36~\AA~$\times$ 7.70~\AA~cell for the $1\bar{1}00$ facet and a hexagonal 6.36~\AA~$\times$ 6.36~\AA~cell for the $0001$ facet. 

The valence electrons considered for each kind of atom were H (1s$^1$), Li (1s$^2$2s$^1$), C (2s$^2$2p$^2$), N (2s$^2$2p$^3$), O (2s$^2$2p$^4$) and S (3s$^2$3p$^4$). The most stable relaxed configurations for the considered solvents in their isolated as well as adsorbed states on electrode surfaces were calculated using the semi-local PBE functionals~\cite{Burke1997,Burke2009} while taking into account van der Waals corrections (DFT-D3 in VASP)~\cite{grimme2010,grimme2011}. The geometry optimizations were performed with an energy cut-off of 520 eV on a plane wave basis by identifying the most stable configuration of each of the solvent molecules after orientating it in all possible vertical and horizontal directions on the surface. All other atoms were allowed to relax until the total free energy was converged to less than 10$^{-6}$ eV.

In order to ensure convergence of the ScGW0 calculations, several convergence tests with respect to k-points, energy cut-off for determining quasiparticle energies, number of empty bands and frequency points were performed. It was found that for the $1\bar{1}00$ facet, 228 empty bands and 96 points on the frequency grid were sufficient for convergence, whereas for the $0001$ facet up to 324 empty bands and at least 144 points on the frequency grid were needed. In both cases an energy cut-off of 150 eV was deemed to be sufficient as going to 175 eV showed very small changes in the estimation of the band structures and edges ($<$ 0.1 eV). For the rectangular $1\bar{1}00$ cell, a 7 $\times$ 6 $\times$ 1 k-point grid yielding 16 points in the Irreducible Brillouin Zone (IBZ) was used, whereas for the hexagonal $0001$ cell, a 7 $\times$ 7 $\times$ 1 k-point grid, yielding 25 points in the IBZ. ScGW0 calculations for the bulk Li$_2$O$_2$ phase were also performed in a cell containing 4 stoichiometric units of Li$_2$O$_2$ using a hexagonal cell of size 3.18~\AA~$\times$ 3.18~\AA~$\times$ 7.70~\AA~ with 10 $\times$ 10 $\times$ 4 gamma centered k-points and 144 empty bands.

In order to explicitly account for effects of only renormalization due to surface, we also calculated the energy levels of the gas phase molecules in the corresponding empty cells with the same periodicity of their physisorbed counterparts. This is different from existing studies, which calculate the isolated gas phase molecules on a 1 $\times$ 1 $\times$ 1 k-point grid without accounting for periodicity~\cite{Garcia:2009aa,Garcia:2011aa,Siegel2016}. This methodology ensures that effects due to periodic solvent interactions are cancelled when estimating the amount of renormalization. Although, it must be noted that the effect of periodicity itself was found to be very negligible when comparing solvents in as molecules in isolated and periodic simulation boxes. Further, the impact of solvation also has been found to be very minor ($<$ 0.1 eV) in a study of very similar systems using an implicit solvation approach~\cite{Siegel2016}, and has been therefore ignored in the present work.  This will form the focus of future investigations.

\section{\label{sec:level1}Results and Discussion}

\begin{figure}[!ht]
\centering
\includegraphics[width=0.48\textwidth]{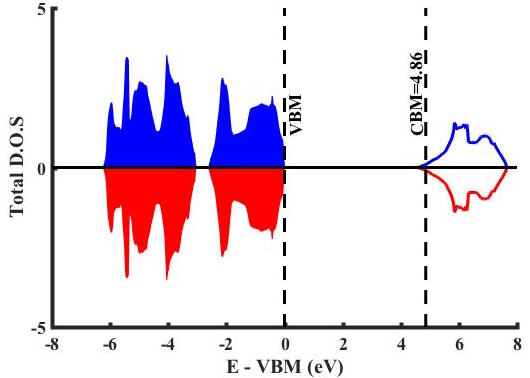}
\caption{The total density of states of the bulk Li$_2$O$_2$ phase.}
\label{fig:bulk_tdos}
\end{figure}

The ScGW0 method is very sensitive to the choice of parameters employed for achieving convergence.  The computations were found to be extremely memory intensive with each case, requiring a minimum of 8 TB of total memory distributed over 120 processors over a period of 24 hours for each update of self-energy and orbitals. ScGW0 calculations on the bulk phase (F{\"o}ppl structure) were performed to obtain a preliminary benchmark of the parameters employed in this work. The total density of states for this system can be seen in Fig.~\ref{fig:bulk_tdos}. The valance band maximum (VBM) and the conduction band minimum (CBM) are also indicated in the plot yielding a band gap of 4.86 eV, which can be expected to increase if a fully self-consistent ScGW algorithm is employed. The value is in good agreement with previous studies~\cite{Hummelshoj:2010aa,Garcia:2011ab,Radin:2012aa}, implying robustness of the method and the convergence parameters used here. 

\begin{figure*}[!ht]
\centering
\includegraphics[width=1.0\textwidth]{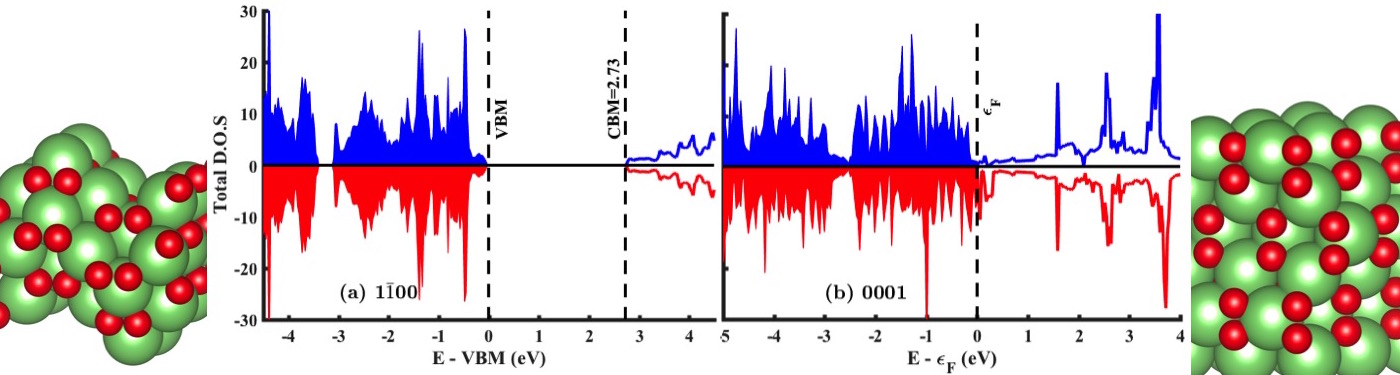}
\caption{The total density of states and the respective geometries for the bare (a) $1\bar{1}00$ facet and (b) $0001$ facet of the Li$_2$O$_2$ cathode surface.}
\label{fig:bare_dos}
\end{figure*}

In Fig.~\ref{fig:bare_dos}, we plot the total density of states (DOS) for the bare $1\bar{1}00$ and $0001$ facets of the Li$_2$O$_2$ cathode surface with each of their geometries considered in the computational model on the side. As can be seen in Fig.~\ref{fig:bare_dos}a, the $1\bar{1}00$ facet is semi-conducting/insulating with a gap of 2.73 eV between the VBM and the CBM. The spin up and spin down channels for this facet are found to be symmetric. On the other hand, the $0001$ facet was found to be half-metallic, as seen by its total DOS in Fig.~\ref{fig:bare_dos}b, which is in agreement with the work of Radin \etal~\cite{Radin:2012aa}. Unlike the $1\bar{1}00$ facet, the $0001$ facet is found to have asymmetric spin-up and spin-down electron channels, with the Fermi level ($\epsilon_{\mathrm{F}}$) located just above the edge of the spin-up channel. The gap between the occupied and unoccupied states in either channel in this work is observed to be smaller than that observed in the work of Radin \etal~\cite{Radin:2012aa}, and we believe that apart from stricter convergence parameters used in this work, the difference arises due to the usage of a stoichiometric slab in this work, in contrast to the symmetric slab used by Radin \etal.~\cite{Radin:2012aa}

The HOMO-LUMO gaps for the considered solvent molecules calculated in the 2 supercells at the PBE and the ScGW0 levels in this work are mentioned in Table~\ref{tab:homo-lumo-ads}. The predicted energy levels and band edges for isolated molecules were found to be in excellent agreement with those from the work of \textit{Siegel}~\etal~\cite{Siegel2016}, who employed a similar methodology and isolated molecules without any periodicity, the effect of which is found to be very miniscule in our work. It also can be seen that using the PBE functional without any corrections leads to severe underestimation of even molecule band gaps.

\begin{table}
\caption{\label{tab:homo-lumo-ads}%
HOMO-LUMO gaps, $\Delta$ (in eV) of solvent molecules calculated with the uncorrected PBE functional and the  ScGW0 corrections in the supercells considered for the $1\bar{1}00$ and $0001$ surface facets.}
\begin{ruledtabular}
\begin{tabular}{lcccc}
\textrm{Solvent}&
\textrm{$\Delta_{\mathrm{PBE}}^{1\bar{1}00}$}&
\textrm{$\Delta_{\mathrm{ScGW0}}^{1\bar{1}00}$}&
\textrm{$\Delta_{\mathrm{PBE}}^{0001}$}&
\textrm{$\Delta_{\mathrm{ScGW0}}^{0001}$}\\
\colrule
ACN & 8.74 & 10.93 & 8.53 & 10.91 \\
DME & 4.66 & 8.13 & 4.5 & 8.02\\
THF & 4.44 & 8.15 & 4.25 & 8.12\\
DMSO & 4.37 & 7.58 & 4.32 & 7.53 \\
\end{tabular}
\end{ruledtabular}
\end{table}

\subsection{\label{sec:charge_surf} Renormalization on the 1$\bar{1}$00 Surface}

\begin{figure*}[!ht]
\centering
\includegraphics[width=1.0\textwidth]{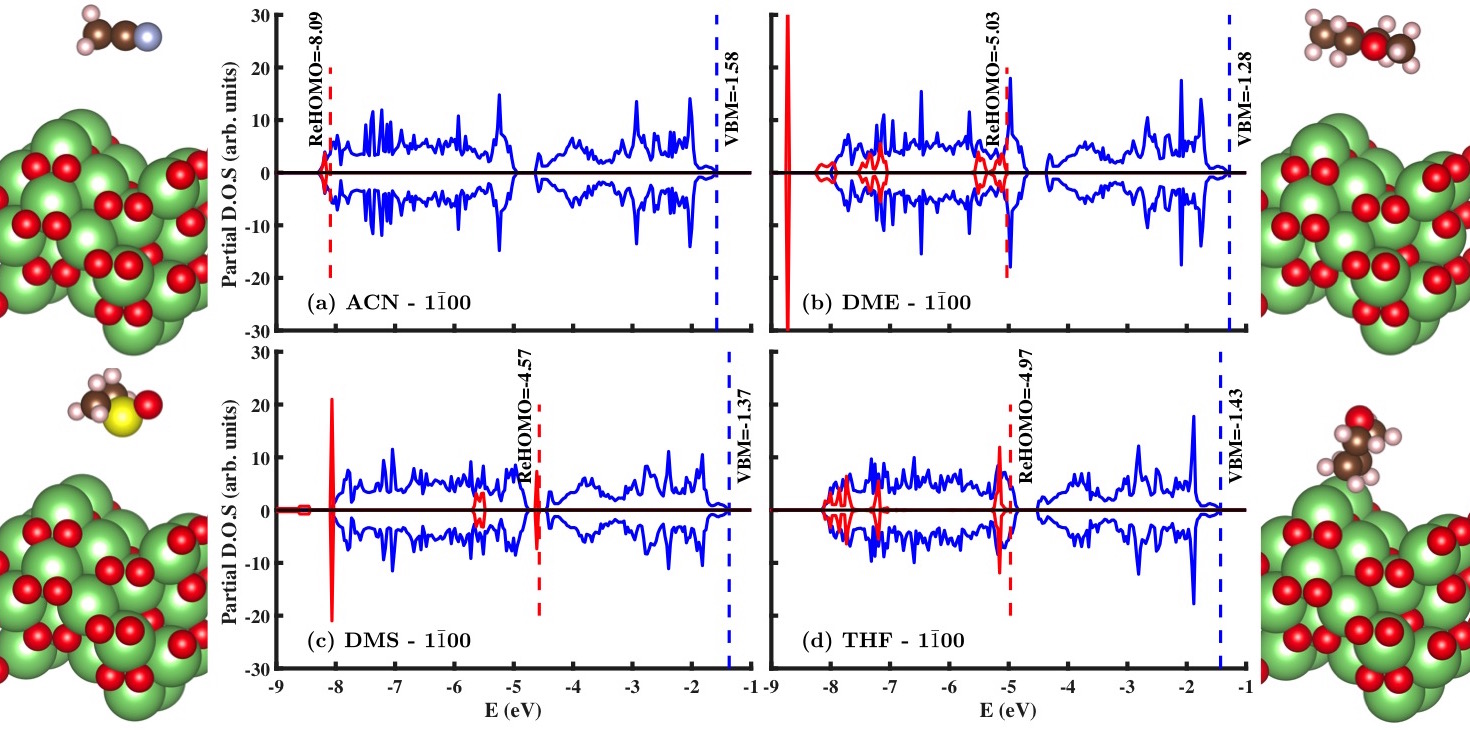}
\caption{Site-projected partial density of states of solvent molecules (red) and the $1\bar{1}00$ terminated Li$_2$O$_2$ surface (blue) for solvents (a) ACN (b) DME (c) THF and (d) DMSO (clockwise) with their converged DFT optimized geometries and indicated locations of frontier energy levels.}
\label{fig:sol1-100}
\end{figure*}

Next, we performed the same set of calculations to examine the renormalization of the HOMO levels of the solvent molecules, and the VBM and CBM of the $1\bar{1}00$ facet of the surface. In Fig.~\ref{fig:sol1-100}, we plot the partial density of states of the $1\bar{1}00$ facet and the four physisorbed solvent molecules along with their converged geometries displayed on the side. The most stable orientation of ACN (a) was found to be parallel to the $1\bar{1}00$ facet with the N-C-C bond aligned to the cell side. The C-C-C-C chain in DME (b) was found to align diagonally in the cell with a vertical dip towards the O-rich trough on the $1\bar{1}00$ facet, possible because of limitations in the lateral cell size. THF (c) favored a vertical orientation with the C-C-C-C ring in proximity to the Li-rich crest on the $1\bar{1}00$ facet, whereas DMSO (d) favored a parallel orientation with the O atom pointing away from the surface.

In Table~\ref{tab:renorm-1-100}, we tabulate the renormalized energy levels of both the $1\bar{1}00$ surface facet (VBM and CBM) and the solvent molecules (ReHOMO and ReLUMO), which were obtained from the analysis of the site projected partial density of states. Also defined is $\delta$ = (VBM - ReHOMO), which is the difference in energy between the renormalized HOMO level of the solvent molecule and the VBM of the surface, and Re$\Delta$ = (ReLUMO - ReHOMO), which is the renormalized HOMO-LUMO gap of physisorbed solvent molecules.

\begin{table}[t]
\caption{\label{tab:renorm-1-100}%
Renormalized energy levels (eV) of the $1\bar{1}00$ surface facet, VBM and CBM, and the solvent molecules, ReHOMO and ReLUMO, along with  $\delta$ = (VBM - ReHOMO) and Re$\Delta$ = (ReLUMO - ReHOMO).}
\begin{ruledtabular}
\begin{tabular}{lcccccc}
\textrm{Solvent}&
\textrm{VBM}&
\textrm{CBM}&
\textrm{ReHOMO}&
\textrm{ReLUMO}&
\textrm{$\delta$}&
\textrm{Re$\Delta$}\\
\colrule
ACN & -1.58 & 1.18 & -8.09 & 1.75 & 6.51 & 9.84\\
DME & -1.28 & 1.42 & -5.03 & 2.17 & 3.75 & 7.20\\ 
THF & -1.43 & 1.48 & -4.97 & 2.05 & 3.54 & 7.02\\
DMSO & -1.37 & 1.33 & -4.57 & 2.02 & 3.20 & 6.59\\
\end{tabular}
\end{ruledtabular}
\end{table}

On comparison with data from Table~\ref{tab:homo-lumo-ads}, it can be observed that the HOMO-LUMO gap shrinking of all solvent molecules is similar, with MeCN showing the maximum shrinking (1.09 eV) and DME showing the minimum shrinking (0.93 eV). Considering the fact that the $1\bar{1}00$ facet is insulating with a gap of 2.73 eV, it is not surprising at all that the HOMO-LUMO gap shrinking across all solvent molecules is of a similar order of magnitude. As shown and explained in the work of \textit{Garcia-Lastra}~\etal~\cite{Garcia:2009aa,Garcia:2011aa}, the renormalization effects get much stronger as the surface gets more metallic. 

In order to first rank the solvents for their oxidative stability, we note that the number of electrons per O$_2$ transferred during discharge or the O$_2$ reduction reaction (e$^-$/ORR) and during charge or the O$_2$ evolution reaction (e$^-$/OER), both given by Eq.~\ref{eq:li2o2}, is a very intuitive and powerful measure of cell performance. Ideally, both the formation and decomposition of Li$_2$O$_2$ should yield 2e$^-$ per O$_2$ molecule. For several common solvents, the e$^-$/ORR was found to be mostly confined to a 2e$^-$/O$_2$ limit within a nominal error limit ($\leq$2~$\%$) with a similar discharge overpotential~\cite{McCloskey:2012aa}, although a nearly 2e$^-$/O$_2$ discharge behavior does not guarantee that all the electrons are involved in the formation of stoichiometric Li$_2$O$_2$~\cite{McCloskey:2013aa}. During charge, however, the deviation of e$^-$/OER from an ideal 2e$^-$/O$_2$ process was found to occur markedly different ways for different solvents~\cite{McCloskey:2012aa}, indicating the presence of undesirable parasitic processes~\cite{McCloskey:2013aa} in varying degrees. Further, one can assert that the larger the deviation, the less efficient and rechargeable the cell is.

In an earlier work, we demonstrated that, to a first approximation, the highest occupied molecular orbital energy level of the solvent in gas phase is a good descriptor for oxidative stability, which is accounted for by choosing e$^-$/OER as the metric of rechargeability~\cite{Khetan:2014aa}. However, the highest occupied molecular orbital (HOMO) and lowest unoccupied molecular orbital (LUMO) levels of the solvent molecule in its gas or even bulk liquid phase are not representative of its true state at the interface where it interacts with the electrode components. To establish a descriptor for oxidative stability based on renormalization effects, we refer back to the thermodynamic model of electron energy alignments developed earlier in Fig.~\ref{fig:new_band_scheme}, and that the VBM of the surface also undergoes renormalization. We surmise that the probability of an electron transfer from the solvent to hole states at the VBM on the surface will depend not on the absolute values of renormalized HOMO level values of solvent molecules, but the difference in the energy levels between the renormalized HOMO level and VBM. This difference has been defined earlier as $\delta$ = (VBM - ReHOMO). Further, during the process of charging, the presence of a charging overpotential $\mathrm{U_{chg}}$ at the cathode surface will lower the Fermi level of the surface by e$\mathrm{U_{chg}}$, thereby reducing the difference between ReHOMO and VBM by the same amount and further increasing the solvent's propensity toward oxidation.

\begin{figure}[!ht]
\centering
\includegraphics[width=0.45\textwidth]{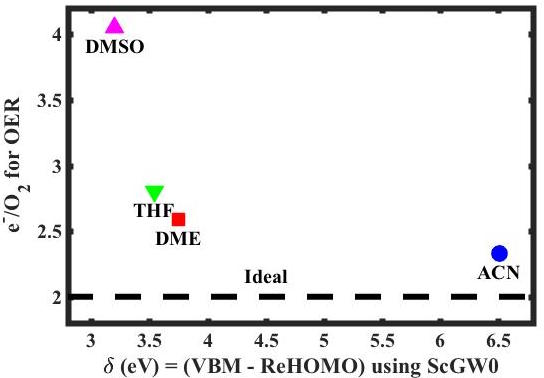}
\caption{Correlation between the chosen metric of rechargeability, e$^-$/O$_2$ for OER, taken from the work of \textit{McCloskey}~\etal~\cite{McCloskey:2012aa} and (a) the calculated difference between the renormalized HOMO level of solvent and VBM energy level of the $1\bar{1}00$ terminated surface, $\delta$}
\label{fig:RenormHOMOcorr}
\end{figure}

To test this hypothesis, we plot our chosen marker of cell performance, e$^-$/OER, adopted from the work of \textit{McCloskey}~\etal~\cite{McCloskey:2012aa} against the calculated values of the difference between the renormalized HOMO level and VBM energy levels, $\delta$, for the four solvents in Fig.~\ref{fig:RenormHOMOcorr}. As can be clearly seen in Fig~\ref{fig:RenormHOMOcorr}a, $\delta$, shows a clear correlation with the rechargeability metric, signaling a greater tendency for electrochemical decomposition of solvents for which this difference is smaller. The results obtained in our work point to an exponential variation of the rechargeability metric with respect to $\delta$, in comparison to a more linear variation with respect to the gas phase HOMO levels~\cite{Khetan:2014aa}.

The analysis performed here leads to a refined picture of solvent-surface interaction and subsequent renormalization of frontier energy levels on the $1\bar{1}00$ facet of Li$_2$O$_2$ during charging. A good descriptor for solvent stability is one which can be computed easily and requires minimal user intervention to calculate. In that sense, the gas phase HOMO level for aprotic solvents in Li-O$_2$ batteries can already be considered a very good descriptor, because it can be easily computed for a large number of solvents which satisfy other requirements from an application point of view. When choosing a descriptor, it is also important to rule out any significant inconsistencies, which in the present case may arise from the first principles method being used for calculation.

\subsection{\label{sec:discharge_surf} Renormalization on the 0001 Surface}

\begin{figure*}[!ht]
\centering
\includegraphics[width=1.0\textwidth]{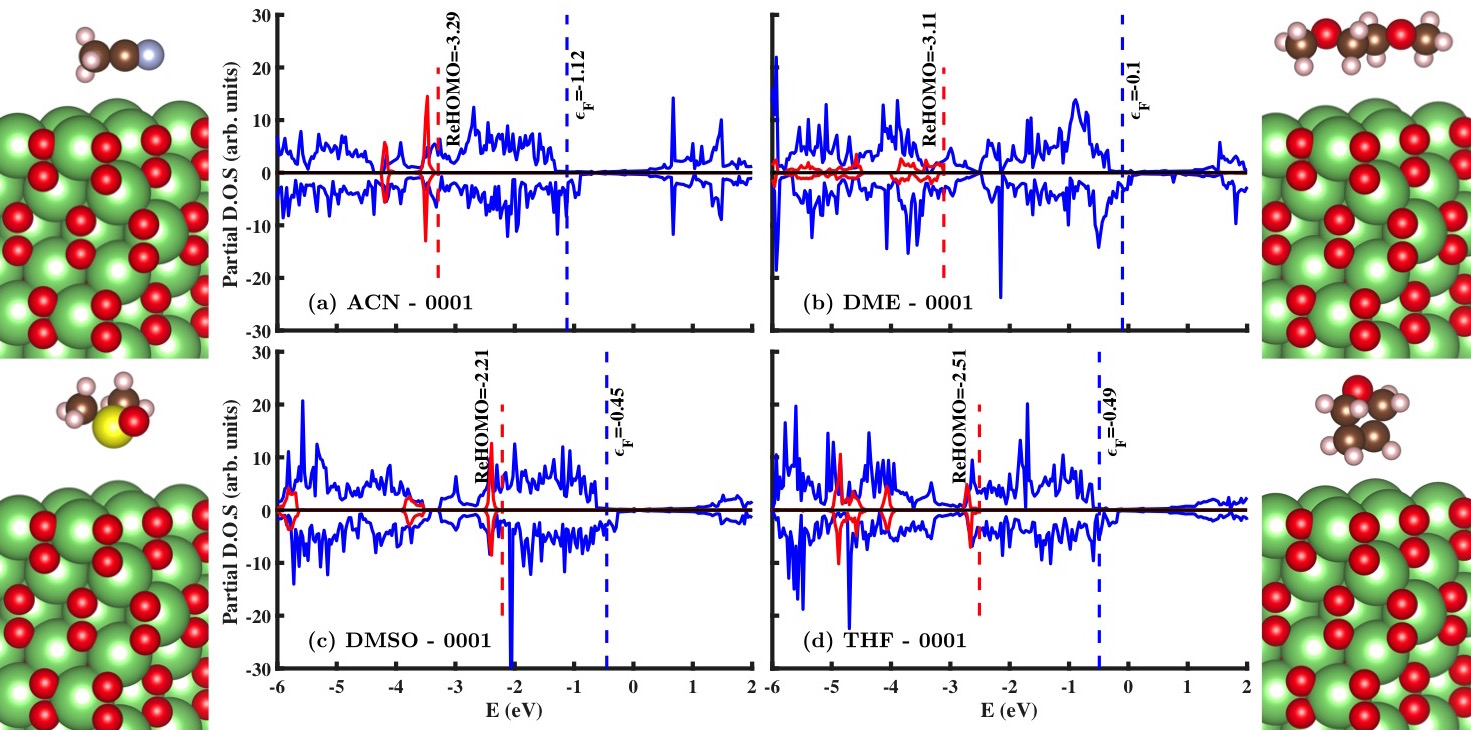}
\caption{Site-projected partial density of states of solvent molecules (red) and the $0001$ terminated Li$_2$O$_2$ surface (blue) for solvents (a) ACN (b) DME (c) THF and (d) DMSO (clockwise) with their converged DFT optimized geometries and indicated locations of frontier energy levels.}
\label{fig:sol0001}
\end{figure*}

The same set of calculations were performed to examine the renormalization of the frontier energy levels of  solvent molecules physisorbed on the metallic $0001$ facet of Li$_2$O$_2$ surface, which is known to be stable under discharging. In Fig.~\ref{fig:sol0001}, we plot the partial density of states of the surface and the four physisorbed solvent molecules along with their converged geometries displayed on the side. The most stable orientations of solvent molecules were found to be identical to those found over the $1\bar{1}00$ facet. Likewise, there was no observed explicit bond formation between the solvent molecule and the surface even upon artificially forcing the solvent molecule close to the surface.

It can be observed that there is asymmetry in the spin-up and spin-down channels of the partial density of states of not only the surface but also the solvent molecules, which we attribute to the method of projection in VASP. It was subsequently verified using Bader charge analysis that there was no charge transfer between the solvent molecules and the surface in any of the cases. 

In Table~\ref{tab:renorm-0001}, we tabulate the renormalized HOMO-LUMO levels of the solvent molecule as well as the Fermi levels, $\epsilon_{\mathrm{F}}$, of the surface, which were obtained from the analysis of the site projected partial density of states, as seen in Fig.~\ref{fig:sol0001}. Also shown is the $\gamma$ = ($\epsilon_{\mathrm{F}}$ - ReHOMO), which is the difference in energy between the renormalized HOMO level of the physisorbed solvent molecule and the Fermi level of the surface, and Re$\Delta$ = (ReLUMO - ReHOMO), which is the renormalized HOMO-LUMO gap of the physisorbed solvent. The $0001$ surface of Li$_2$O$_2$ is metallic, and not surprisingly, the amount of HOMO-LUMO gap shrinking on this surface is found to be much larger as compared to the insulating $1\bar{1}00$ surface. The gap of ACN shrinks by 3.92 eV while the gap of DMSO shrinks by 3.16 eV. The large renormalization can be attributed to high degree of image charge creation due to the surface being metallic in nature, as explained in the work of \textit{Garcia-Lastra}~\etal~\cite{Garcia:2009aa,Garcia:2011aa}.

\begin{table}[t]
\caption{\label{tab:renorm-0001}%
Renormalized Fermi level (eV) of the $0001$ surface facet, $\epsilon_{\mathrm{F}}$, and the HOMO-LUMO levels of the solvent molecules, ReHOMO and ReLUMO, along with  $\gamma$ = ($\epsilon_{\mathrm{F}}$ - ReHOMO) and Re$\Delta$ = (ReLUMO - ReHOMO).}
\begin{ruledtabular}
\begin{tabular}{lcccccc}
\textrm{Solvent}&
\textrm{$\epsilon_{\mathrm{F}}$}&
\textrm{ReHOMO}&
\textrm{ReLUMO}&
\textrm{$\gamma$}&
\textrm{Re$\Delta$}\\
\colrule
ACN & -1.12 & -3.29 & 3.70 & 2.17 & 6.99\\
DME & -0.10 & -3.11 & 1.51 & 3.01 & 4.62\\ 
THF & -0.49 & -2.53 & 2.32 & 2.04 & 4.85\\
DMSO & -0.45 & -2.20 & 2.17 & 1.75 & 4.37\\
\end{tabular}
\end{ruledtabular}
\end{table}

To understand the effect of renormalization on the propensity of solvent oxidation on the $0001$ surface during discharging, it must be first observed that the presence of a discharging overpotential $\mathrm{U_{dis}}$ at the cathode surface will increase the Fermi level of the surface by e$\mathrm{U_{dis}}$, thus increasing $\gamma$  = ($\epsilon_{\mathrm{F}}$ - ReHOMO) by the same amount and ultimately lowering propensity of solvent oxidation for all solvents. Additionally, aprotic solvents are known to degrade predominantly via H-abstraction by means of nucleophilic attack/substitution by the strong nucleophile O$_2^-$ during charging, although there is clear evidence of simultaneous solvent oxidation leading to carbonate formation as well~\cite{McCloskey:2012ac,McCloskey:2013aa,Khetan:2014ab}.

The value of $\gamma$ is comparably smaller for every physisorbed molecule than the corresponding value of $\delta$ on the $1\bar{1}00$ surface, which indicates that solvents may perhaps simultaneously decompose via oxidation during discharging. However, a direct correlation with the metric e$^-$/O$_2$ for ORR is infeasible to make because the values for both all solvents is very close to the ideal value of 2e$^-$/O$_2$ consumed. As noted earlier, a nearly 2e$^-$/O$_2$ discharge behavior does not guarantee that all the electrons are involved in the formation of stoichiometric Li$_2$O$_2$~\cite{McCloskey:2013aa}. Using the metric $\gamma$, we observe that DME, instead of ACN, is the most resistant to degradation via electrochemical oxidation, whereas DMSO is the least resistant. These observations corroborate very well with the measured yield of Li$_2$O$_2$ from the work of \textit{McCloskey}~\etal~\cite{McCloskey:2013aa}, where they demonstrated that that using DME exhibits the highest and DMSO the lowest Li$_2$O$_2$ yield of all solvents across several varieties of used carbon cathodes.

\subsection{\label{sec:defects} Effect of Surface Defects on Renormalization}

In addition to the pristine surfaces, we also considered the effect of defects on the pristine surfaces which are encountered as intermediate states of Li$_2$O$_2$ by the interacting solvent molecules during discharging or charging. The first kind of defect considered was in form of a missing Li atom, henceforth simply called as a $\textit{Li-defect}$ state. This choice was motivated by the fact that the surface formation mechanism of Li$_2$O$_2$ has been proposed in several studies to comprise of two basic lithiation steps~\cite{Viswanathan:2013aa}, the first step being

\begin{equation} \label{eq:lio2*}
\mathrm{Li{^+}} + \mathrm{O{_2}} + \mathrm{e{^-}} \leftrightharpoons \mathrm{LiO{_2^*}}
\end{equation}

and the second step leading to formation of a stoichiometric adsorbed molecule as

\begin{equation} \label{eq:li2o2*}
\mathrm{Li{^+}} + \mathrm{LiO{_2^*}} + \mathrm{e{^-}} \leftrightharpoons \mathrm{Li{_2}O{_2^*}}
\end{equation}

The Li-defect state of the surface represents its intermediate conditions after the first lithiation step. The second kind of defect considered in the analysis was in form of a missing Li$_2$O$_2$ molecule from the pristine surfaces, henceforth called simply as the $\textit{Li$_2$O$_2$-defect}$ state. Such a state, although still stoichiometric, effectively results in a step defect for both terminations~\cite{Hummelshoj:2013aa}. For the present analysis we considered the renormalization of only the ACN and DMSO molecules as they represent the two limiting cases of stability, polarizability of the considered solvent molecules.

\begin{figure*}[!ht]
\centering
\includegraphics[width=1.0\textwidth]{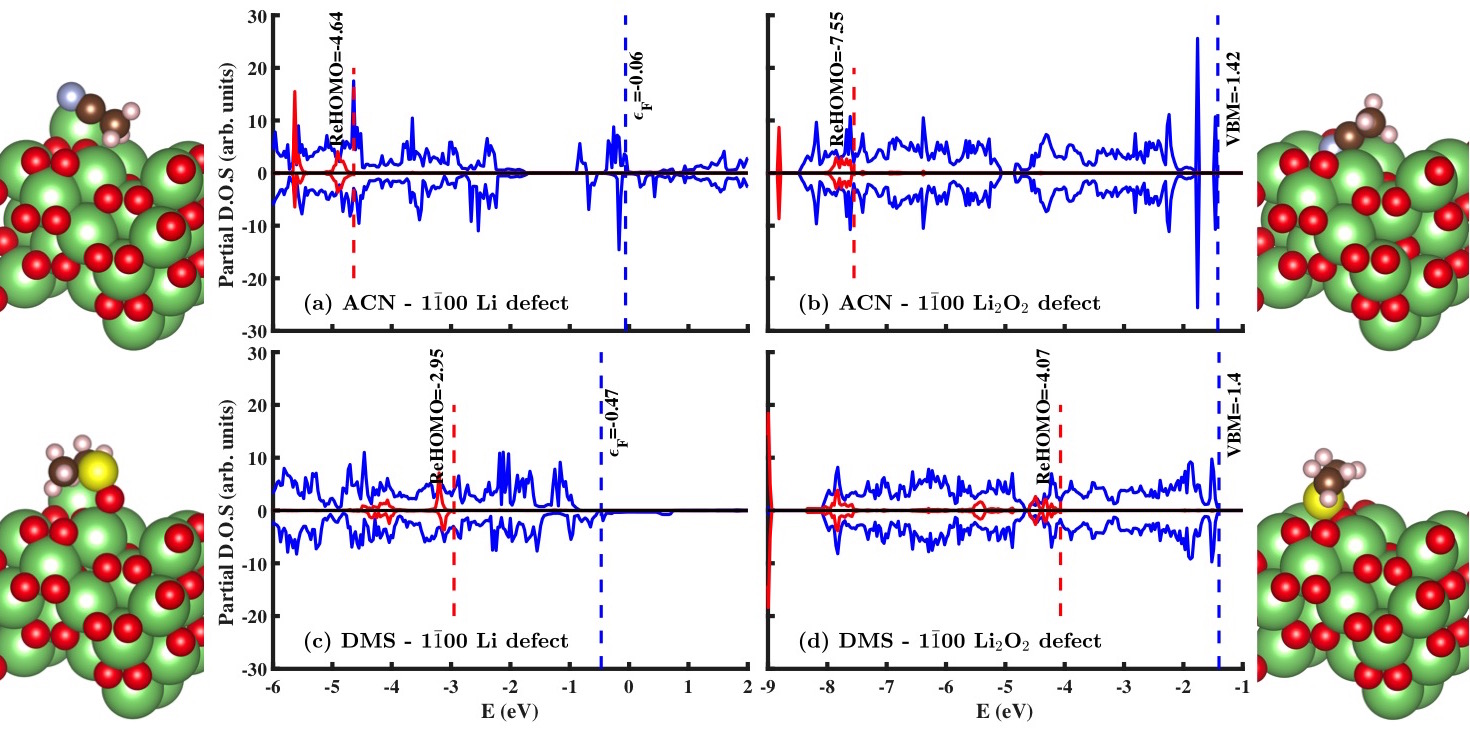}
\caption{Site-projected partial density of states of ACN (red) molecule over the (a) Li defect (blue) and (b) Li$_2$O$_2$ defect (blue) states of the $1\bar{1}00$ surface and of DMSO (red) molecule over the (c) Li defect (blue) and (d) Li$_2$O$_2$ defect (blue) states of the $1\bar{1}00$ surface with their converged DFT optimized geometries and indicated locations of frontier energy levels.}
\label{fig:def1-100}
\end{figure*}

In Fig.~\ref{fig:def1-100}, we plot the partial density of states of the of ACN and DMSO over the Li- and Li$_2$O$_2$-defect states of the $1\bar{1}00$ facet along with their converged geometries displayed on the side. The behavior of the defect surfaces was found to be in stark contrast to that of the pristine surface. The electronic structure of the surface itself was found to be influenced by the nature of the defect. While the Li defect made the surface metallic in the spin-down channel (Fig.~\ref{fig:def1-100}a and Fig.~\ref{fig:def1-100}c), the stoichiometric Li$_2$O$_2$ defect retained the insulating nature of the pristine $1\bar{1}00$ surface (Fig.~\ref{fig:def1-100}b and Fig.~\ref{fig:def1-100}d). Given the electrochemical formation and decomposition of Li$_2$O$_2$ are kinetically quite fast~\cite{Hummelshoj:2013aa}, the intermediate surfaces of Li$_2$O$_2$ possibly undergo rapid changes in their electronic and band structure during the during the charging process. Consequently, these rapidly changes will in turn induce rapid renormalization of the HOMO-LUMO levels of the interacting solvent molecules.

As in the case of pristine surfaces, several orientations of the ACN and DMSO molecules were calculated to find the minimum energy configurations over each of the defect surfaces. The ACN molecule was found to move closer to the surface with Li defect, with the N atom pointing away from the defect site and, in contrast, it displayed adsorption via the N atom on the surface with Li$_2$O$_2$ defect, as opposed to mere physisorption on the pristine $1\bar{1}00$ surface. The effect of defects on the surfaces led to an increased renormalization of the ACN molecule's HOMO-LUMO levels with respect to the surface. The Li defect surface resulted in a value of $\gamma$ = 4.56 eV and the Li$_2$O$_2$ defect surface resulted in a value of $\delta$ = 6.13 eV, as compared to the value of $\delta$ =  6.51 eV on the pristine $1\bar{1}00$ surface.

The DMSO molecule was found to adsorb via the O atom on both the defect surfaces, as opposed to the pristine $1\bar{1}00$ surface, where the most stable physisorbed orientation of the molecule occurred with the O atom pointing away from the pristine $1\bar{1}00$ surface. Like the case of ACN, the Li defect surface ultimately resulted in a value of $\gamma$ = 2.48 eV and the Li$_2$O$_2$ defect surface resulted in a value of $\delta$ = 2.67 eV, as compared to the value of $\delta$ =  3.20 eV on the pristine $1\bar{1}00$ surface.

From the observations above, and the fact that the presence of a charging overpotential will serve to lower the Fermi level or VBM of the surface thereby decreasing $\delta$, the presence of defects will lead to adsorption of solvent molecules and drastically increase their propensity toward oxidation. It can also be clearly ascertained that for both solvent molecules the effect of the metallic Li defect on renormalization is much stronger as compared to the insulating Li$_2$O$_2$ defect surface. It must be noted that not possible defects can be considered in this work owing to limitations on the system size which can be treated with reasonable computational capacity. Further, although the decrease in $\delta$ due to Li defect is larger in case of ACN, such effects do not change any stability trends.

\begin{figure*}[!ht]
\centering
\includegraphics[width=1.0\textwidth]{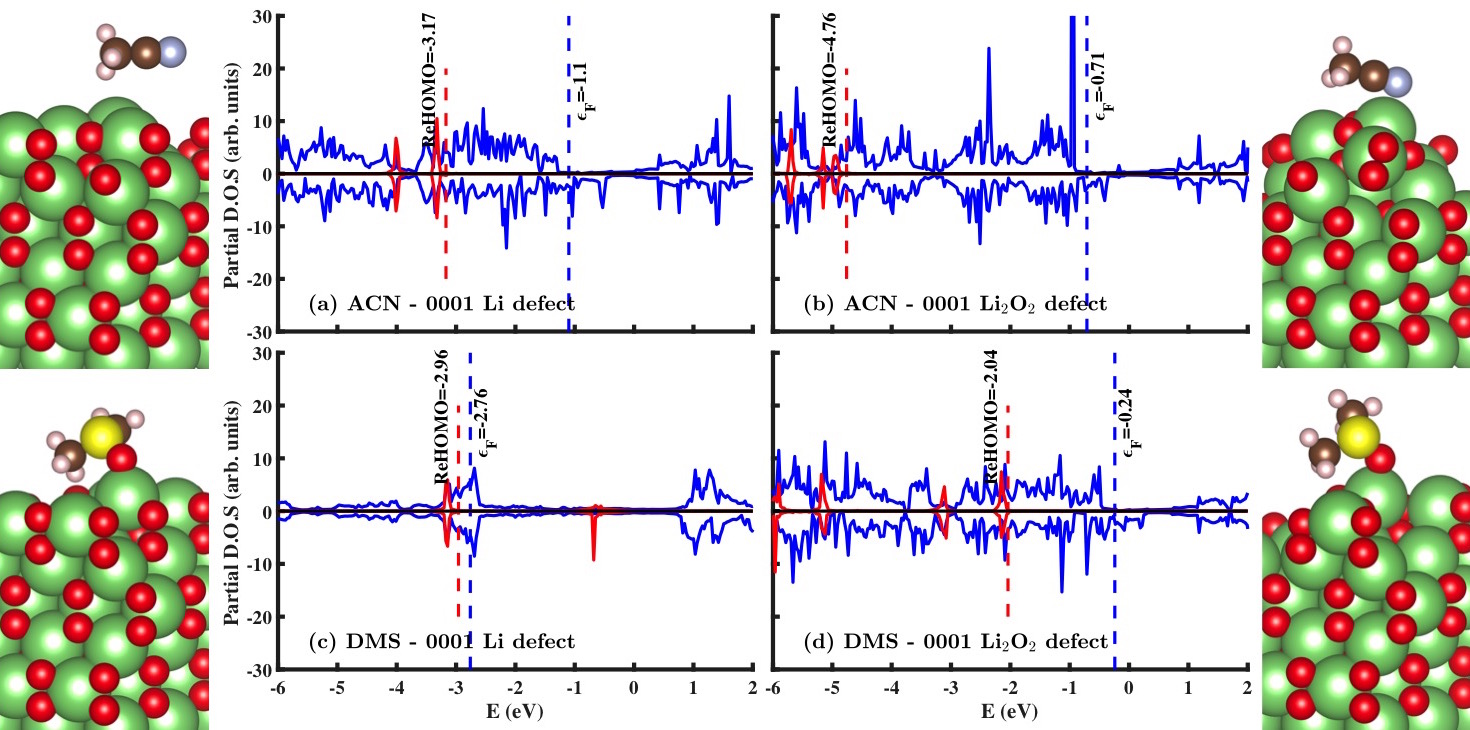}
\caption{Site-projected partial density of states of ACN (red) molecule over the (a) Li defect (blue) and (b) Li$_2$O$_2$ defect (blue) states of the $0001$ surface and of DMSO (red) molecule over the (c) Li defect (blue) and (d) Li$_2$O$_2$ defect (blue) states of the $0001$ surface with their converged DFT optimized geometries and indicated locations of frontier energy levels.}
\label{fig:def0001}
\end{figure*}

Next, we plot the partial density of states of ACN and DMSO over the Li- and Li$_2$O$_2$-defect states of the $0001$ facet along with their converged geometries displayed on the side in Fig.~\ref{fig:def0001}. Although the presence of either kind of defect did not alter the metallic nature of the pristine $0001$ surface, the polarization and subsequent renormalization of the solvent molecules was nevertheless found to be much stronger in this case. Similar to the case of defects on the $1\bar{1}00$ termination, the molecules were found to undergo favorable adsorption in all cases except that of the ACN molecule over the Li defect $0001$ surface. In case of the Li$_2$O$_2$ defect surface, both solvent molecules were observed to cause drastic reorganization of surface atoms resulting in a distortion of the surface structure.

Similar to the case of the $1\bar{1}00$ termination, the effect of Li defects on the $0001$ surface led to an increased renormalization of both solvents' HOMO-LUMO levels with respect to the surface. The Li defect surface resulted in a value of $\gamma$ = 2.07 eV as compared to a value of $\gamma$ =  2.17 eV for ACN and in a value of $\gamma$ = 0.2 eV as compared to a value of $\gamma$ =  1.75 eV for DMSO. It must be emphasized that certain trial orientations of the DMSO molecule resulted in its complete disintegration and splitting of the Li defect  $0001$ surface, thus demonstrating a very high degree of instability. Remarkably, in the same system the two spin channels were found to be almost symmetric. The effect of Li$_2$O$_2$ defects on the $0001$ surface led to varying degrees of renormalization of both solvents' HOMO-LUMO levels with respect to the surface. The Li$_2$O$_2$ defect surface ultimately resulted in a value of $\gamma$ = 4.05 eV as compared to a value of $\gamma$ =  2.17 eV for ACN and in a value of $\gamma$ = 1.8 eV as compared to a value of $\gamma$ =  1.75 eV for DMSO. 

While the presence of a discharge overpotential will increase the value of $\gamma$ between the solvents and the the $0001$ defect surface, it is clear from the above analysis that a sizable proportion of solvent degradation during discharging could occur via electrochemical charge transfer. In either process, charging or discharging, the effects of defects on the surfaces can considerably alter the dynamic polarization of both the surface and the adsorbate. Other kinds of defects on pristine surfaces, like kinks and terraces, will induce a different degree of polarization and accompanying renormalization of the solvents' HOMO-LUMO levels. The thermodynamic driving force associated with the adsorption energy of the solvent molecules on such surfaces will ultimately play a crucial role in determining their stability against oxidation. 

\section{\label{sec:level1}Summary and Conclusions}

In this work, we addressed in detail the influence of interfacial interactions on the electrochemical stability of nonaqueous solvents during charging and discharging by providing an accurate description of the molecular energy levels of the solvents (HOMO-LUMO) as well as the electrode surfaces (VBM-CBM) using high fidelity GW calculations. Additionally, we employed a new methodology of using the difference between the renormalized HOMO level and VBM energy levels instead of the absolute renormalized HOMO levels of the solvent molecule, and demonstrated its robustness and straightforward applicability, as one need not align the energy levels of all systems to a computed vacuum for the purpose of subsequent comparison.

The HOMO-LUMO gap of a solvent molecule was found to shrink by $\approx$ 1 eV when the molecules were physisorbed on the $1\bar{1}00$ surface, which is known to be stable during charging. It was shown that on the insulating $1\bar{1}00$ facet of the Li$_2$O$_2$ cathode surface, the phenomenon of renormalization does not affect our previous hypothesis of choosing the gas phase HOMO level of solvent molecules as the descriptor of their oxidative stability. Using a first order descriptor, such as the gas phase HOMO level, to rank solvents for their oxidative stability would be a more efficient method for initial screening of a large number of solvents. High accuracy parameters and convergence checks were used for computing the renormalized energy levels and such a procedure is computationally very intensive, and is therefore not recommended for a first order screening of several thousands of molecules. The renormalization of solvents' HOMO-LUMO levels was found to be much more evident over the metallic $0001$ surface, with DME instead of ACN, emerging as the most stable solvent molecule against oxidation.

The complete analysis of the defect surfaces along with the solvents ACN and DMSO points to three important observations. Firstly, solvent oxidation is most likely to occur at defect sites rather than on pristine surfaces. This is evidenced from the lowering $\delta$ and $\gamma$ values, as well as the actual adsorption of solvent molecules on the defect surfaces. Secondly, the polarization of the defect surfaces will be very dependent on its stoichiometry as well as on the interacting solvent molecule, which may or may not undergo adsorption. Lastly and most importantly, the trends in solvent stability observed in the pristine surfaces will still hold when the same surface is considered as the surface for polarization. The absence of solvent-salt interactions as well as the consideration of other surfaces of Li$_2$O$_2$ can turn out to be of significance, and this will be the subject of future studies.

\begin{acknowledgments}
A. K. acknowledges the funding for his doctoral studies by the Deutsche Forschungsgemeinschaft (DFG). A. K. also acknowledges the grant of computational time for the project by the J{\"u}lich Aachen Research Alliance (JARA) via Project Number - 99000538(JHPC3601). A.K. and V. V. acknowledge support from Convergent Aeronautics Solutions (CAS) project under the NASA Aeronautics Research Mission Directorate.  V. V. also acknowledges support from the National Science Foundation CAREER award CBET-1554273.
\end{acknowledgments}

\bibliography{bibsources}

\end{document}